\documentstyle[12pt,epsfig]{article}
\def\lapp{{\ \lower 0.6ex \hbox{$\buildrel<\over\sim$}\ }}
\def\gapp{{\ \lower 0.6ex \hbox{$\buildrel>\over\sim$}\ }}
\begin{document}
\begin{titlepage}
\vspace*{-1cm}
\begin{flushright}
DTP/95/88  \\
October 1995 \\
\end{flushright}
\vskip 1.cm
\begin{center}
{\Large\bf Fermiophobic Higgs bosons at the Tevatron   \\ [3mm]}
\vskip 1.cm
{\large A.G.~Akeroyd\footnote{A.G.Akeroyd@durham.ac.uk} }
\vskip .4cm
{\it Department of Mathematical Sciences, Centre for Particle
Theory,\\ University of Durham, \\
Durham DH1 3LE, England. }\\
\vskip 1cm
\end{center}

\begin{abstract}

Higgs bosons with negligible couplings to fermions can
arise in various non--minimal Higgs sectors. We show that such a particle
could be discovered during the current run at the Tevatron,
and would be evidence against a minimal supersymmetric Higgs sector.

\end{abstract}
\end{titlepage}
\newpage
%%%%%%%%%%%%%%%%%%%%%%%%%%%%%%%%%%%%%%%%%%%%%%%%%%%%%%%%%%%%%%%%%%
\section{Introduction}
The Standard Model (SM) \cite{Wein} has proved remarkably successful to date in
describing the particle interactions of nature. However, the theory requires
that the electroweak symmetry is broken and an efficient way of accomplishing
this is to introduce scalar particles (Higgs bosons) with non--zero
vacuum expectation values (VEVs) \cite{Higgs}. Thus far no such particles
have been
detected and therefore it is prudent to explore all possible Higgs sectors.
The minimal SM consists of one complex isospin Higgs doublet which after
symmetry breaking predicts one physical neutral scalar ($\phi^0$), although
much can be found in the literature concerning extended models \cite{Gun}
i.e. the
non--minimal SM\footnote{Defined by assuming no other new particles apart
from Higgs bosons.}. Extended Higgs sectors with additional doublets/triplets
always require exotic Higgs bosons with electric charge ($H^{\pm}$)
and zero tree--level couplings to gauge bosons ($A^0$). Also
possible in some extended models is `fermiophobia' \cite{Hab}, \cite{Pois}
i.e. zero tree--level
couplings to fermions. Such particles ($H_F$) can only arise in certain
Higgs models, and in particular are {\sl not} predicted by the minimal
supersymmetric model (MSSM). Therefore the discovery of a $H_F$ would be
evidence against a minimal supersymmetric extended Higgs sector.

Our work is organised as follows. In Section 2 we
describe the various Higgs models which can contain fermiophobia, and then
investigate the properties of $H_F$.
Section 3 deals with the phenomenology of $H_F$ at both the Tevatron and
proposed Tevatron upgrade. Finally Section 4 contains our conclusions.

\section{Models with Fermiophobia}
The most theoretically favourable non--minimal Higgs sectors are those that
contain only doublet representations. These naturally
keep $\rho\equiv M^2_W/(M^2_Z\cos^2\theta_W)\approx 1$ \cite{Ross}.
 Models with triplets
can also be considered and the most popular of these was proposed by
Georgi and Machacek containing one--doublet and two--triplets \cite{Geo},
\cite{Veg}, \cite{Bam}. In this paper
we shall consider the various two--Higgs--doublet models (2HDM) of which there
are four distinct versions \cite{Bar}, and the above mentioned
Higgs triplet model (HTM).
\begin{table}[htb]
\centering
\begin{tabular} {|c|c|c|c|c|} \hline
 & Model~I & Model~I$'$ & Model~II & Model~II$'$  \\ \hline
u (up--type quarks)   & 2 & 2 & 2 & 2 \\ \hline
d (down--type quarks) & 2 & 2 & 1 & 1 \\ \hline
e (charged leptons)   & 2 & 1 & 1 & 2 \\ \hline
\end{tabular}
\caption{The four distinct structures of the 2HDM.}
\end{table}

Table 1 shows the four different ways with which the 2HDM can be coupled to
the fermions. The numbers (1 or 2) show which Higgs doublet couples to
which fermion type. Natural flavour conservation \cite{Wein1}
requires that at most one doublet can couple to any
particular fermion--type. Model~II is the structure required for the
MSSM \cite{Gun}, \cite{Fay} and thus it has received substantially
more attention in the literature.
Model~I is the only model that can display fermiophobia and this becomes
clear when we view the couplings in Table 2.
We are interested here in the lighter of the two neutral, CP--even Higgs
bosons ($h$).
\begin{table}[htb]
\centering
\begin{tabular} {|c|c|c|c|c|} \hline
& Model~I & Model~I$'$ & Model~II & Model~II$'$  \\ \hline
$hu\overline u$ & $\cos\alpha/\sin\beta$ & $\cos\alpha/\sin\beta$
& $\cos\alpha/\sin\beta$  & $\cos\alpha/\sin\beta$      \\ \hline
$hd\overline d$ & $\cos\alpha/\sin\beta$& $\cos\alpha/\sin\beta$
&$-\sin\alpha/\cos\beta$ &$-\sin\alpha/\cos\beta$      \\ \hline
$he\overline e$ &  $\cos\alpha/\sin\beta$  & $-\sin\alpha/\cos\beta$
  &$-\sin\alpha/\cos\beta$
&  $\cos\alpha/\sin\beta$\\ \hline
\end{tabular}
\caption{The fermion couplings of $h$ in the 2HDM relative to those for the
 minimal SM Higgs boson ($\phi^0$).}
\end{table}

Here $\alpha$ is a mixing angle used to diagonalize the CP--even mass matrix
and $\beta$ is defined by $\tan\beta=v_2/v_1$ ($v_i$ is the VEV of the $i^{th}
$ doublet and $v_1^2+v_2^2=246$ GeV$^2$).
 From Table 2 we see that fermiophobia is only possible in Model~I
if $\cos\alpha\to 0$ \cite{Hab}. We note that the heavier CP--even Higgs
($H$) in Model~I
would itself be fermiophobic if $\cos\alpha\to 1$. However this particle could
 be substantially heavier than $h$ and so is not considered. From now on we
 shall label the fermiophobic Higgs in this model as being $h$, with $H_F$
 referring to any generic fermiophobic Higgs.
Therefore it is apparent that fermiophobia is not possible
in the MSSM since it requires Model II type couplings. Hence searching for
$H_F$ is well motivated. We note that another signal of the 2HDM (Model I)
 which is not possible in the MSSM would be the discovery of a light
 $H^{\pm}$ ($M_{H^{\pm}}\le
M_W$); this is possible through direct pair production at LEP2 \cite{Ake1}
or top quark decay at the Tevatron \cite{Ake3}.

The other model that we shall study and contains fermiophobia is the HTM.
Predicted here are two fermiophobic neutral bosons, $H_5^0$ and $H_1^{0'}$.
In this paper we shall not consider a charged $H_F$; for
recent studies of the latter we refer the reader to Refs.~\cite{Cheung},
\cite{God}. In an earlier paper \cite{Ake2} we analysed this model using a
natural
argument of equating all Higgs self couplings ($\lambda_i$) to 1;
it was shown that $H_1^{0'}$ can be
taken as a physical mass eigenstate and we also obtained the following natural
 mass hierarchy (with $v^2=246$ GeV$^2$):
\begin{equation}
M^2_{\psi_1}=10v^2\to
16v^2\;,\;\;M^2_{H_5}=3v^2\;,\;\;M^2_{H_3}=v^2\;,\;\;\;M^2_{\psi_2}=0\to
1.5v^2\;.
\end{equation}
The compositions of the mass eigenstates $\psi_1$ and $\psi_2$ are given by
\begin{eqnarray}
\psi_1=H^{0'}_1\sin\alpha_T+H^0_1\cos\alpha_T\;, \\
\psi_2=H^{0'}_1\cos\alpha_T-H^0_1\sin\alpha_T\;,
\end{eqnarray}
with $H^0_1$ being a neutral scalar similar to that of the minimal SM, and
$\alpha_T$ being a mixing angle.
Ref. \cite{Ake2} shows that $\sin\alpha_T\le 0.05$ or $0.999\le
\sin\alpha_T\le 1$,
and so negligible mixing occurs in Eqs. (2) and (3). Therefore $H_1^{0'}$
could be the lightest or the heaviest of the bosons in the HTM
depending on the exact value of the angle $\alpha_T$. We shall be
concentrating on the scenario of it being the lightest but will also mention
detection prospects if this is not the case. Ref. \cite{Ake2} constrains
$\alpha_T$ by using the bound
$\sin\theta_H\le 0.63$,\footnote
{$\sin\theta_H$ is the analogy of $\tan\beta$ for the HTM, defined by
$\sin\theta_H=\sqrt{8b^2}/\sqrt {a^2+8b^2}$, with the doublet (triplet) VEV
denoted by a (b).}
 found from considering the effects of $H^{\pm}_3$ on
the $Z\to b\overline b$ vertex. This result is for $M_{H_3}\le 200$ GeV which
we see as being justified if one wishes to search for $H_1^{0'}$ at the
Tevatron; from the ratios in Eq. (1) we see that $M_{H_3}\le 200$ GeV would
imply $0\le M_{\psi_2} \le 245$ GeV, which is the mass range that is relevant
at
the Tevatron. The other neutral fermiophobic Higgs, $H^0_5$, is likely to be
heavier than $H_1^{0'}$ (if $\psi_2\equiv H_1^{0'}$) and we shall see that it
is harder to produce at the Tevatron due its more suppressed couplings to
vector bosons. In Ref.~ \cite{Ake2} we proposed that the detection of a $H_F$
would suggest the HTM, since the 2HDM (Model I) requires fine--tuning for
fermiophobia.

\begin{figure}[p]
\begin{center}
\mbox{\epsfig{file=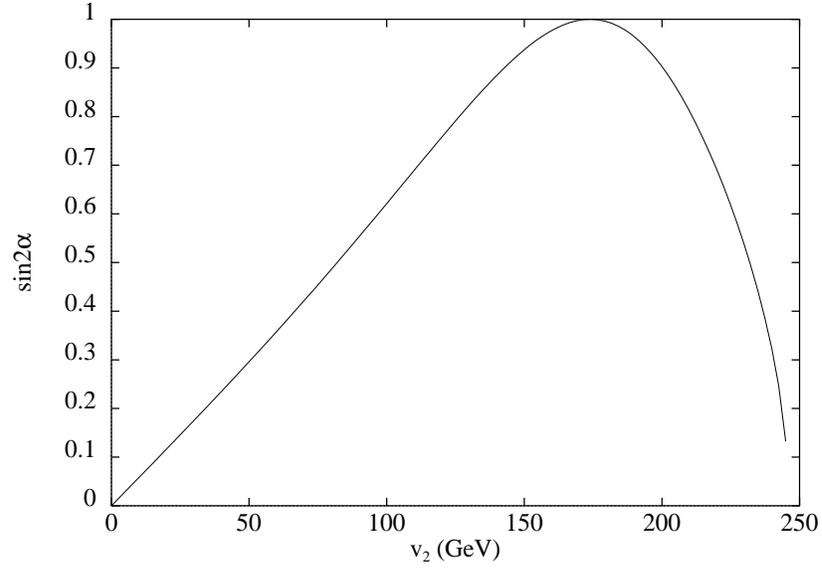,angle=-90,height=8.5cm}}
\end{center}
\vspace{-5mm}
\caption{$\sin2\alpha$ as a function of $v_2$.}
\label{Fig:fig1}
\end{figure}

It is possible to apply the above natural argument to the 2HDM (Model~I)
to see the variation of $\sin\alpha$ with $\tan\beta$. Plotted in Figure
\ref{Fig:fig1} is $\sin2\alpha$ as a function of $v_2$. We see that maximal
mixing
($\sin2\alpha=1$, $\alpha=45^{\circ}$) occurs when $v_2=v_1\approx 174$ GeV.
For $v_2\gg 174$ GeV (i.e. $\tan\beta\gg 1$), the two $\alpha$ solutions for
$\sin2\alpha$ approach $0^{\circ}$ and $90^{\circ}$. Hence for fermiophobia
($\alpha\to 90^{\circ}$) this argument would require larger $\tan\beta$, a
result consistent with the bound $\tan\beta\ge 1.25$ for $M_{H^\pm}\le 200$
GeV \cite{Gross}.

\begin{figure}[p]
\begin{center}
\mbox{\epsfig{file=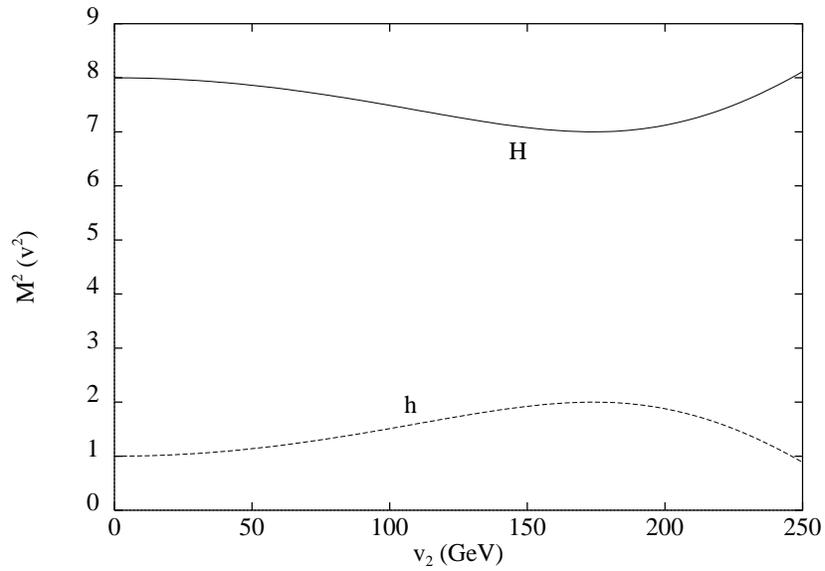,angle=-90,height=8.5cm}}
\end{center}
\vspace{-5mm}
\caption{The squared masses of $H$ and $h$ as a function of $v_2$}
\label{Fig:fig2}
\end{figure}

We may also obtain the analogous mass hierarchy (see Eq. (1)) for the 2HDM
(Model~I). Figure \ref{Fig:fig2} shows the squared masses of $H$ and $h$ as
a function of $v_2$, and from this we see that $7v^2\le M^2_H \le 8 v^2$ and
$v^2\le M^2_h\le 2v^2$. Therefore we have
\begin{equation}
M^2_H=7v^2\to
8v^2\;,\;\;M^2_h=v^2\to 2v^2\;,\;\;M^2_{H^{\pm}}=v^2\;,\;\;\;M^2_{A^0}=v^2\;.
\end{equation}
Eq. (4) suggests that $h$ is likely to be of comparable mass to
$M_{H^{\pm}}$ and so
justifies the use of the bound $\tan\beta\ge 1.25$ for $M_{H^{\pm}}\le 200$
GeV,
if one wishes to search for $h$ at the Tevatron.

\begin{figure}[hp]
\begin{center}
\mbox{\mbox{\epsfig{file=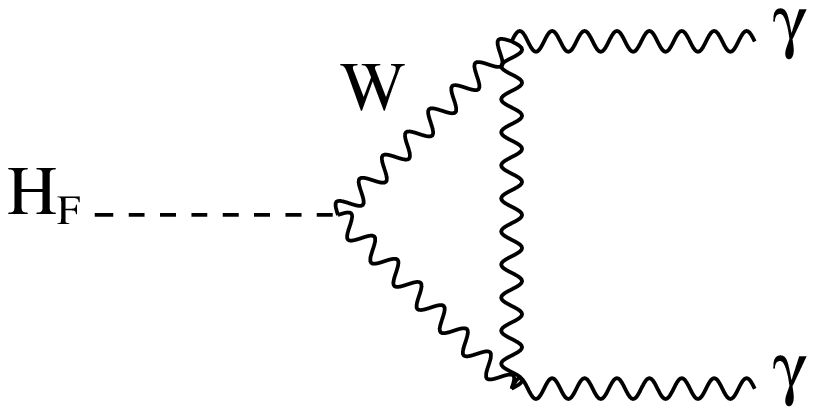,height=3cm}}
\mbox{\epsfig{file=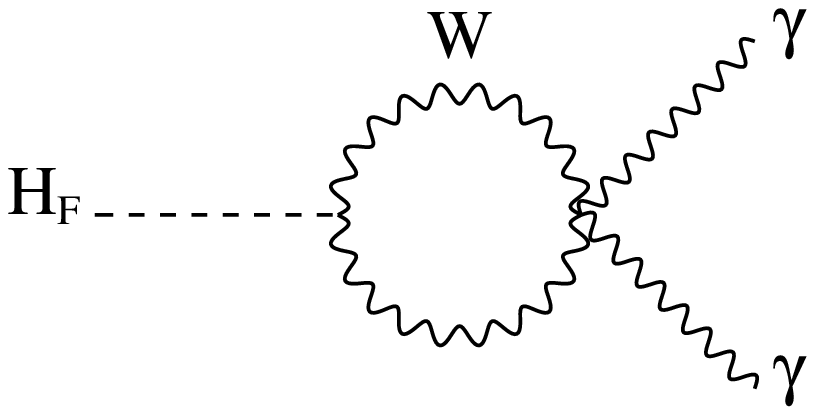,height=3cm}}}
\end{center}
\vspace{-5mm}
\caption{Two--photon decay of $H_F$.}
\label{Fig:ferm1}
\end{figure}

We now study the branching ratios (BRs) of $H_F$. Tree--level decays to
fermions are obviously not allowed, and if $M_{H_F}\le 80$ GeV then the
only possible
tree--level channels are $H_F\to W^*W^*$, $Z^*Z^*$, with `*' denoting an
off--shell vector boson\footnote{Not including
decays to other Higgs bosons which will be
heavily off--shell also.}. Since these latter decays are not very strong
(the vector bosons being considerably off--shell)
then one--loop mediated decays can compete and these are displayed in Figure
\ref{Fig:ferm1}.
For the case of $H_F\to \gamma\gamma$, the $W$ mediated decays give the
dominant contribution \cite{Gun}, \cite{Ell} and only these are included .
The one--loop decays to $f\overline f$
are renormalization scheme dependent and it is conventional in the
literature to consider an extreme fermiophobic Higgs with the renormalized
$H_F\to f\overline f$ vertex set equal to
zero \cite{Stan}, \cite{Diaz}. The BRs predicted by Refs. \cite{Stan} and
\cite{Diaz} agree and imply that the channel $H_F\to \gamma\gamma$ dominates
for $M_{H_F}\le 80$ GeV; at
$M_{H_F}\approx 95$ GeV the tree--level process $H_F\to WW^*$ is equally
likely as $H_F\to \gamma\gamma$, each having BR=$45\%$. In contrast, for
$\phi^0$ and the lightest neutral CP--even scalar of the MSSM the branching
ratio to two photons is of the order $0.1\%$. For higher $M_{H_F}$
the vector boson channels dominate along with decays to other Higgs bosons
($H_F\to t\overline t$ is not allowed at tree--level). Therefore the
distinctive fermiophobic signature of $H_F\to \gamma\gamma$ is disappearing
 for
$M_{H_F}\ge 100$ GeV, and so we shall focus on the region of $M_{H_F}\le 100$
GeV. For the heavier mass region ($\ge 160$ GeV) the only difference between
the decays of $H_F$ and $\phi^0$ would be due to the presence of lighter
Higgs bosons e.g. $h\to A^0Z$.

The BRs used in Ref. \cite{Stan} are for a $H_F$ with $\phi^0$ strength
(i.e. minimal SM strength) couplings to vector bosons. This is not the case
for the $H_F$ that we are considering, as can be seen from Eqs. ($5\to 7$).
The couplings here are
expressed relative to those of the minimal SM Higgs boson (with $s_H\equiv
\sin\theta_H$) \cite{Veg}:
\begin{equation}
H^{0'}_1W^+W^-: {2\sqrt 2\over {\sqrt 3}}s_H\;,\;\;\;\;\;
 H^{0'}_1ZZ:{2\sqrt 2\over \sqrt 3}s_H\;,
\end{equation}
\begin{equation}
H^0_5W^+W^-: {1\over {\sqrt 3}}s_H\;,\;\;\;\;\;H^0_5ZZ:{-2\over {\sqrt 3}}
s_H\;,
\end{equation}
\begin{equation}
hW^+W^-:-\cos\beta\;,\;\;\;\;\;hZZ:-\cos\beta\;.
\end{equation}

Eqs. (5) and (7) show that both the $H_FW^+W^-$ and $H_FZZ$
 couplings for $H^{0'}_1$ and $h$ are scaled by the same amount, and so the
 BRs used in Ref. \cite{Stan} can be used. This is not true for $H^0_5$
 which has an enhanced $H^0_5ZZ$ compared to $H^0_5W^+W^-$. However, in
 the region of $M_F\le 80$ GeV (which is of interest to us)
 the channel $H^0_5\to Z^*Z^*$ is small and so we may use the results in
 Ref. \cite{Stan} to a very good approximation.

\section{Phenomenology at the Tevatron}
\begin{figure}[htp]
\begin{center}
\mbox{
\mbox{\epsfig{file=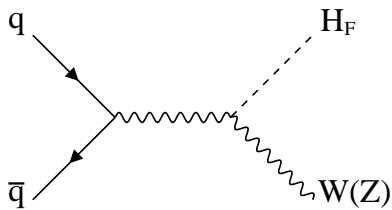,height=3cm}}
}
\end{center}
\vspace{-5mm}
\caption{The main production mechanism of $H_F$ at the Tevatron.}
\label{Fig:ferm3}
\end{figure}

For $\phi^0$ the main production process at the Tevatron ($\sqrt s=1.8$ TeV)
is gluon--gluon fusion via a top quark loop \cite{Geo2}. This is not
allowed for $H_F$ and
nor are any diagrams involving associated production with top quarks \cite{Ng},
\cite{Kun}.
Therefore
there remains two processes; associated production with vector bosons
\cite{Glas}
 and
vector boson fusion \cite{Cahn}. However, Ref. \cite{Stan} shows
that the latter gives less events and so we shall
focus on the former whose Feynman diagram is displayed in Figure
\ref{Fig:ferm3}.
As mentioned in Section 2, Ref. \cite{Stan} assumed minimal SM strength
couplings to vector bosons for $H_F$ and so the production cross sections for
$H^{0'}_1$, $H^0_5$ and $h$ relative to those for $\phi^0$  will scale
by the squares of the couplings given in Eqs. ($5\to 7$). Thus for the
process $q\overline q\to W^*\to WH_F$ we have the following cross section
ratios:
\begin{equation}
H^0_5:H^{0'}_1:h:\phi^0={1\over 3}s^2_H:
{8\over 3}s^2_H:\cos^2\beta:1\;, \end{equation}
and for $q\overline q\to Z^*\to ZH_F$
\begin{equation}
H^0_5:H^{0'}_1:h:\phi^0={4\over 3}s^2_H:
{8\over 3}s^2_H:\cos^2\beta:1\;. \end{equation}
Due to the bounds $\sin^2\theta_H\le 0.39$ and $\cos^2\beta\le 0.39$ we see
that $H^{0'}_1$ may be produced with $\phi^0$ strength in both channels,
while $h$ has at best a cross section 0.39 that of $\phi^0$. $H^0_5$ has
very weak couplings to $W^+W^-$ (at best 0.13 that of $\phi^0 W^+W^-$), but
better to $ZZ$.

We note that a $H_F$ with $\phi^0$ strength couplings to $ZZ$ would have been
seen at LEP if $M_{H_F}\le 60$ GeV \cite{Aleph}. Eq. (9) shows that this
lower bound will in general be weaker for $H^{0'}_1$, $h$ and $H^0_5$.
The method of searching for $H_F$ at the Tevatron is described in Ref.
\cite{Stan}
and we shall briefly review it here. The photons from $H_F$ act as a trigger
for the events, and then various cuts are applied depending on whether the
vector bosons decay hadronically or leptonically. For the leptonic decay
it is shown that the main background
($W\gamma\gamma$ and $Z\gamma\gamma$) is negligible. Hence we only require
a reasonable number of events ($\ge 3$) in this channel for detection. For the
hadronic decays of the vector bosons there is a background
($jj\gamma\gamma$).\footnote {From processes like $gg$, $qq\to
qq\gamma\gamma$.}
For this channel the $WH_F$ and $ZH_F$ signals are combined due to the
invariant mass distribution being unable to separate the $W$ and $Z$ peaks.
Table 3
shows the expected number of signal and background events\footnote
{The event numbers in all our tables are obtained
from Ref. \cite{Stan} with appropriate scaling for a particular Higgs model.}
for 67 pb
$^{-1}$ of data, which is the current data sample at the Tevatron. The numbers
are for $H^{0'}_1$ with $s_H^2=0.39$, its maximum value.
\begin{table}[htb]
\centering
\begin{tabular} {|c|c|c|c|} \hline
$M_{H_F}$ (GeV) & $WH/ZH$ (leptonic) & $WH/ZH$ (jets) & $jj\gamma\gamma$
\\ \hline
 60 & 9.8/7.7& 50.9 & 3.5 \\ \hline
 80& 3.7/3.5 &  20.2 & 1.9  \\ \hline
100 & 0.6/0.5 & 3.1 &  1.0 \\ \hline
\end{tabular}
\caption{Number of signal and background events for the process
$q\overline q\to W^*(Z^*) \to H^{0'}_1 W(Z)$, with $H^{0'}_1\to \gamma\gamma$
and $W\to l\nu$, $Z\to l\overline l$, $\nu\overline \nu$, or $W$, $Z\to jj$.}
\end{table}

 From Table 3 we see that the region $M_{H_F}\le 80$ GeV can be covered
with $\ge 3$ events in the background free leptonic channel, and a
$\ge 4.3\sigma$ signal in the hadronic channel. With 140 pb$^{-1}$ available
by the end of 1995 the event numbers in Table 3 will be increased by a factor
of approximately 2.1. This would enable the region $M_{H_F}\le 90$ GeV to be
covered, i.e. the mass at which the $\gamma\gamma$ decay starts to fall
rapidly.
It is very possible that $s_H^2$ is considerably less than 0.39, and if
this is the case then the signal becomes weaker. With 140 pb$^{-1}$ of
luminosity and $M_{H_F}=60$ (80) GeV  one can obtain $\ge 3$
events in the leptonic channel if $s_H^2\ge 0.06$
($s^2_H\ge 0.16$).

For the case of $h$ the maximum number of signal events is less due to
the cross section being proportional to $\cos^2\beta$. Table 4 is the
analogy of Table 3 for $h$ with $\cos^2\beta=0.39$.
\begin{table}[htb]
\centering
\begin{tabular} {|c|c|c|c|} \hline
$M_{H_F}$ (GeV) & $WH/ZH$ (leptonic) & $WH/ZH$ (jets) & $jj\gamma\gamma$
\\ \hline
 60 & 3.7/2.9 & 19.1 & 3.5 \\ \hline
 80& 1.4/1.3 & 7.6 & 1.9  \\ \hline
100 & 0.2/0.2 & 1.2 &  1.0 \\ \hline
\end{tabular}
\caption{Same as for Table 3 but for the process $q\overline q\to W^*
(Z^*)\to h W(Z)$.}
\end{table}

We see that $M_F\le60$ GeV can be probed ($\ge 3$ events in the leptonic
channel and a $\ge 4\sigma$ signal in the hadronic channel). The coverage
increases to $M_F\le 80$ GeV with 140 pb$^{-1}$.

Would it be possible to distinguish between $H^{1'}_0$ and $h$ ? If $s_H^2$
is near its maximum of $0.39$ then the cross section for a given $M_F$ is
considerably larger for $H^{1'}_0$ than that for $h$
(see Eqs. (8) and (9)). Once the mass of $H_F$ is measured one can estimate
the cross section and thus distinguish between the two models. Of course
a sufficient number of $\gamma\gamma$ events will be needed to
measure the mass and so one should use the hadronic channel.
Sufficient events should be present, certainly up to $M_F\approx 80$ GeV.

The above analysis has assumed that the lighter mass eigenstate $\psi_2$
is composed dominantly of $H^{1'}_0$ (see Eq. (3)). If this is not the case
then $\psi_2\approx H^0_1$ and the heavier eigenstate $\psi_1$ will be
equal to $H^{1'}_0$. Therefore $H^0_5$ will be the lighter $H_F$ in the HTM.
If $M_{H_5^0}\le 90$ GeV then one may search for the $\gamma\gamma$ decays
at the Tevatron, and the mass hierarchy (Eq. (1)) would suggest that $H^0_3$,
$H^{\pm}_3$, and $H^0_1$ would also be light. However, Ref. \cite{Stan}
shows that at least 1000 pb$^{-1}$ of luminosity would be needed to search
for the SM Higgs ($\phi^0$), and so more would be needed for $H^0_1$ which has
$\phi^0$ strength couplings only in the limit of $s_H\to 0$. The three--plet
bosons ($H^0_3$ and $H^{\pm}_3$) would be difficult to detect at the Tevatron
and prospects are much better at LEP2. The doubly charged Higgs
($H^{\pm\pm}_5$) is likely to have a similar mass to $H^0_5$ (the five--plet
members are degenerate at tree--level) and would offer the best signature of
the HTM. Returning to $H^0_5$, we find that if $s_H^2=0.39$ then $\ge 3$ events
are predicted in the $Z^*\to ZH^0_5$ leptonic channel for $M_{H_F}\le 60$
(80) GeV with
a data sample of 67 pb$^{-1}$ (140 pb$^{-1}$). Prospects for detection are
therefore approximately the same as for $h$.

It is probable that the Tevatron will be upgraded in luminosity with 2
fb$^{-1}$ being possible by the year 2000. The increased number of events would
allow heavier $M_{H_F}$ to be probed. For $H^{0'}_1$ with $s^2_H=0.39$ one
would expect $\ge 3$ events in the leptonic channel if $M_{H_F}\le 110$ GeV.
To probe beyond this mass region requires another large increase in luminosity
due to the rapid weakening of BR ($H_F\to \gamma\gamma$). In Ref. \cite{Ake2}
we suggested that the theoretical motivation for the HTM would require $s_H\ge
0.1$ ($s^2_H\ge 0.01$). For this `minimum' value the upgraded Tevatron would
produce $\ge 3$ events in the leptonic channel if $M_{H_F}\le 80$ GeV.
Therefore the coverage would be superior to that of LEP2, the latter only
being able to probe the region $M_{H_F}\le \sqrt s-100$ GeV if $H_F$ has
$\phi^0$ strength couplings. For previous searches at LEP see Refs.~
\cite{Aleph}, \cite{Rizz}.

\section{Conclusions}
We have studied the detection prospects of fermiophobic Higgs bosons ($H_F$)
at the Fermilab Tevatron. Such particles do not possess a tree--level coupling
to fermions and can arise in various non--minimal Higgs models. Importantly,
fermiophobia is not possible in the minimal supersymmetric model (MSSM)
and thus
searching for $H_F$ is well motivated. We considered the 2HDM (Model I) and
the HTM in which can arise the fermiophobic bosons $H^{0'}_1$, $h$ and $H^0_5$.
The dominant decay channel for $M_{H_F}\le 80$ GeV is $H_F\to
\gamma\gamma$, and backgrounds are small. Such a decay has a branching ratio
 of the order $~0.1\%$
for the minimal SM Higgs ($\phi^0$) and the lightest CP--even
Higgs of the MSSM ($h^{SUSY}$). If the $H_FVV$ ($V=W$ or $Z$)
coupling is close to its maximum value then with 140 pb$^{-1}$ of data at the
Tevatron a strong signal would be present for $H^{0'}_1$ ($h$, $H^0_5$) if
$M_{H_F}\le 90$ (80) GeV. It is possible to distinguish $H^{0'}_1$ from
$h$ and $H^0_5$ due to the possibility of a significantly larger cross section,
although we suggested that the mere detection of a $H_F$ would indicate the
HTM.

Prospects are improved at an upgraded Tevatron (2 fb$^{-1}$). For $H^{0'}_1$
with maximum $H_FVV$ coupling detection is possible if $M_{H_F}\le 110$ GeV.
This collider covers more parameter ($M_{H_F}$, $s_H$) space than is possible
at LEP2. For larger $M_F$ the decay channel $H_F\to \gamma\gamma$ weakens
rapidly and thus the distinctive signature of $H_F$ becomes increasingly
difficult to extract. For this higher mass region, it might be possible to
distinguish $H_F$ from $\phi^0$ and $h^{SUSY}$ due to the absence of
$b\overline b$ decays; this branching ratio is significant for
$\phi^0$ and $h^{SUSY}$ if $M_{\phi^0}\le 150$ GeV but negligible for $H_F$
($H_F\to WW^*$ dominates).
For still heavier $H_F$ (which would be in the range of
the Large Hadron Collider), distinguishing would require the
observation of decays of $H_F$ to lighter Higgs bosons e.g. $H_F\to
A^0Z$, which would not be present for $\phi^0$. However, the latter decays
are also possible for $h^{SUSY}$.
If the Tevatron is not upgraded then the Large Hadron Collider should
cover the range inaccessible at LEP2 (i.e. $M_{H_F}\ge 80$ GeV).
Studies of detection prospects in the $\gamma\gamma$ channel for $\phi^0$
at this collider have been performed \cite{Proc}. The conclusion is that
detection is possible if a very high di-photon mass resolution can be
achieved; this is partly due to BR~ ($\phi^0\to \gamma\gamma)\approx 0.1\%$.
For $H_F$ the significantly larger
BR~ $(H_F\to \gamma\gamma$) would make detection much easier as long
as the production cross section is not too suppressed relative to $\phi^0$.

\section*{Acknowledgements}
I wish to thank W.J. Stirling and S. Willenbrock for useful comments.
This work has been supported by the UK EPSRC.

\end{document}